\documentclass[a4paper]{jpconf}
\usepackage{amsmath}
\usepackage{url}
\usepackage{csquotes,graphicx}
\usepackage{mhchem,booktabs}
\usepackage[utf8]{inputenc}
\usepackage[T1]{fontenc}
\usepackage{mathtools}

\begin{document}

\title{Constraint Analysis for the Interaction of the Vector-Meson Octet with the Baryon Octet}

\author{Y. \"Unal, A. K\"u\c{c}\"ukarslan, S. Scherer}

\address{Institut f\"ur Kernphysik, Johannes Gutenberg-Universit\"at, D-55099 Mainz, Germany,\\
\c{C}anakkale Onsekiz Mart University, Department of Physics, 17100 \c{C}anakkale, Turkey}

\ead{uenal@kph.uni-mainz.de, yaseminunal@comu.edu.tr}

\begin{abstract}
We describe a constraint analysis for the interaction of the vector-meson octet with the baryon octet.
   Applying Dirac's Hamiltonian method, we verify that the standard interaction in terms of two independent
SU(3) structures is consistent at the classical level.
   We argue how the requirement of self consistency with respect to perturbative
renormalizability may lead to relations among the renormalized coupling constants of the system.
\end{abstract}

\section{Introduction}
   In quantum chromodynamics (QCD), the physics of strongly interacting particles is described in terms
of quarks and gluons as the dynamical degrees of freedom (DOF).
   In the low-energy regime, QCD can be approximated by an effective
field theory (EFT) using hadronic dynamical DOF.
   To that end, one writes down the most general Lagrangian consistent
with the assumed symmetries of the fundamental
theory~\cite{Weinberg:1978kz}.
   In addition, one needs a power-counting scheme for the expansion of
physical observables.

   The interaction of the pseudoscalar octet $(\pi,K,\eta)$ with the baryon
octet is largely constrained by spontaneous symmetry breaking \cite{Scherer:2002tk}.
   It is also interesting to investigate the coupling of the
vector-meson octet to the baryon octet.
   The situation is more complex, because a Lorentz-invariant description of spin-1
systems introduces unphysical degrees of freedom.
   Therefore, one imposes constraints which, for an interacting theory,
may lead to relations among the coupling constants of the Lagrangian.

   We investigate the lowest-order effective Lagrangian for the interaction of
the vector-meson octet with the baryon octet by performing a classical Dirac
constraint analysis~\cite{Dirac:2001}.
   For the quantized theory we demand that it is perturbatively renormalizable
in the sense of effective field theory \cite{Weinberg:mt}.
   For the pure vector-meson sector, such an investigation results in a massive
SU(3) Yang-Mills theory~\cite{Djukanovic:2010tb,Neiser:2011}.
   For the interaction of the vector mesons with the baryons we search
for additional relations among the coupling constants.

\section{Classical Dirac Constraint Analysis}
   The most general effective Lagrangian for a system of a massive vector-meson octet
interacting with a massive baryon octet can be written as
\begin{equation}
\label{lagrangian}
\begin{aligned}
\mathcal{L}=&\; \mathcal{L}_\text{1}+\mathcal{L}_\text{1/2} + \mathcal{L}_{\text{int}}+\cdots,\\
\mathcal{L}_{1}  =& -\frac{1}{4} V_{a \mu\nu}  V_a^{\mu\nu}
+ \frac{M_{V}^2}{2} V_{a\mu} V_a^{\mu}
- gf_{abc}\partial_{\mu}V_{a\nu} V_b^{\mu} V_c^{\nu}- \frac{g^2}{4} f_{abc}f_{ade}V_{b\mu}V_{c\nu}V_d^{\mu} V_e^{\nu},\\
\mathcal{L}_\text{1/2}=&\; \bar{\Psi}_{a}\left(i\gamma^\mu \partial_\mu-m\right)\Psi_{a},\\
\mathcal{L}_\text{int}=&\; i \text{G}_\text{F}f_{abc} \bar{\Psi}_{a}\gamma^{\mu}\Psi_{b} V_{c\mu} +\text{G}_\text{D}d_{abc}  \bar{\Psi}_{a}\gamma^{\mu}\Psi_{b}V_{c\mu},
\end{aligned}
    \end{equation}
where the SU(3) indices $a, b, c$ range from $1$ to $8$. The three coupling constants $g$,
$\text{G}_\text{F}$, and $\text{G}_\text{D}$ are dimensionless.
   In Eq.~(\ref{lagrangian}), the ellipses stand for "nonrenormalizable" higher-order
interactions, which we assume to be suppressed by powers of some large scale, as well as for interactions
with other hadrons.
   We take the Lagrangian to be invariant under \emph{global} SU(3) transformations.
   For the vector-meson self interaction, the constraint analysis of Refs.~\cite{Neiser:2011,Bijnens:2014fya}
has already been incorporated, leading to a reduction from five independent couplings
to one single coupling $g$.

In the canonical formalism, the momentum field variables conjugate to the field variables are given by
    \begin{equation}
\pi_{a\mu}  =\frac{\partial \mathcal{L}}{\partial {\dot{V}_a^{\mu}}}=
\frac{\partial \mathcal{L}_1}{\partial {\dot{V}_a^{\mu}}}, \quad \quad
\Pi_{\Psi a}  = \frac{\partial ^R\mathcal{L}}{\partial \dot{\Psi}_{a}}=
\frac{\partial ^R\mathcal{L}_\text{1/2}}{\partial \dot{\Psi}_{a}}, \quad \quad
\Pi_{\Psi^\dagger a} = \frac{\partial ^L\mathcal{L}}{\partial \dot{\Psi}_{a}^\dagger}
=\frac{\partial ^L\mathcal{L}_\text{1/2}}{\partial \dot{\Psi}_{a}^\dagger}.
    \end{equation}
   Because $\psi$ and $\psi^{\dagger}$ are anti-commuting variables, we consider the convention of taking fermionic derivatives
from the left for the adjoint field and from the right for the field.
   Using these relations, we immediately see that the velocities cannot be expressed in terms of the momenta.
   In this case, we cannot go from the configuration space of the system to its phase space.
   To define the Hamiltonian of the system, we introduce three so-called primary constraints \cite{Dirac:2001},
\begin{equation}
{\theta}_{Va}^1 =\pi_{a0} + gf_{abc}V_{b0}V_{c0}\approx0,
\quad \quad \chi_{\Psi a}  = \Pi_{\Psi a}-\frac{i}{2}\Psi_{a}^\dagger\approx0,
\quad \quad    \chi_{\Psi^\dagger a}= \Pi_{\Psi^\dagger a}+\frac{i}{2}\Psi_{a} \approx0.
\label{p}
\end{equation}
   Here, a relation such as ${\theta}_{Va}^1 \approx0$ denotes a weak equation in Dirac's sense \cite{Dirac:2001},
namely that one must not use one of these constraints before working out a Poisson bracket.
   In total, Eq.~(\ref{p}) amounts to 8 constraints for the vector mesons and $8\cdot 4+8\cdot 4=64$ constraints
for the baryons.
   We introduce unknown Lagrange multiplier functions $(\lambda_{a}, \lambda^\dagger_a, \lambda_{Va})$,
define a constraint Hamiltonian (density) $\mathcal{H}_\text{c}$, and construct the total Hamiltonian (density)
in terms of a Legendre transformation,
\begin{equation}
\mathcal{H}_\text{c}=\lambda_{a}\chi_{\Psi a}+\lambda_{a}^{\dagger} \chi_{\Psi^{\dagger} a} +\lambda_{V a}\theta_{Va}^{1}, \quad\quad
\mathcal{H}_{T}=\mathcal{H}_{1} + \mathcal{H}_\text{1/2} + \mathcal{H}_\text{int}+\mathcal{H}_\text{c}.
\end{equation}
\begin{center}
    \begin{table}[ht]
\caption{\label{opt} Counting the DOF for the free vector, Dirac, and interacting theories, respectively.}
\centering
    \begin{tabular}{l*{6}{c}r}
\br
Case   & Total DOF & Constraints & Physical DOF \\
\mr
Free vector fields& 64 & 16 & 48   \\
Free Dirac fields& 128 & 64 & 64   \\
Interacting theory& 192 & 80 & 112   \\
\br
    \end{tabular}
    \end{table}
    \end{center}

   The requirement that Eqs.~(\ref{p}) have to be zero throughout all time results in
\begin{align}
\label{poisson1}
\left\{\theta_{Va}^1, H_{T}\right\}&=i\text{G}_\text{F}f_{abc} \Psi_{b}^{\dagger}\Psi_{c}
    - \text{G}_\text{D}d_{abc} \Psi_{b}^{\dagger} \Psi_{c}-\partial _{i}\pi^{i}_a - M_{V}^2 V_{a0}\nonumber\\
    &\quad +gf_{abc}\pi_{bi}V^i_{c}-gf_{abc}\partial_iV^i_{b}V_{c0}-gf_{abc}\partial_iV_{b0}V^i_{c}\nonumber\\
    &=\theta_{Va}^2 \approx 0, \\
\label{poisson2}
\{\chi_{\Psi a}, H_T\}&=i\partial _{i}\Psi_a^{\dagger}\gamma^0\gamma^i + m\Psi_a^{\dagger}\gamma^0-...-i \lambda_a^{\dagger}=0,\\
\label{poisson3}
\{\chi_{\Psi_a^{\dagger}}, H_T\}&=-i\partial _{i}\Psi_a \gamma^0\gamma^i + m\Psi_a \gamma^0+...-i \lambda_{a}=0,
\end{align}
   where $H_T=\int d^3 x\, {\cal H}_T$.
   Equation (\ref{poisson1}) is a so-called secondary constraint, and, therefore, we obtain 8 additional constraints.
   Evaluating the Poisson bracket of $\theta_{Va}^2$ and $H_T$ results in an equation for the Lagrange multiplier
$\lambda_{Va}$.
   From Eqs.~(\ref{poisson2}) and (\ref{poisson3}) we can solve for the Lagrange multipliers $\lambda^\dagger_a$
and $\lambda_{a}$, respectively.
   At this stage, we have solved for all the Lagrange multipliers and have generated the correct number of
constraints.
   As a result of Dirac's constraint analysis, at the classical level we have a self-consistent theory
with the correct number of constraints and thus the correct number of physical DOF without any
relation among the couplings.

\section{Perturbative Renormalizability}
   Pertubative renormalizability in the sense of EFT requires that the ultraviolet divergences of loop diagrams
can be absorbed in the redefinition of the masses, coupling constants, and fields of the {\it most general} effective Lagrangian.
   As in Refs.~\cite{Djukanovic:2010tb,Neiser:2011,Bijnens:2014fya,Djukanovic:2004mm},
we expect additional relations among the coupling constants due to the perturbative renormalizability condition.
   Redefining the fields and parameters, the counter-term Lagrangian reads
    \begin{equation}
\begin{aligned}
\mathcal{L}_\text{ct}=&
-\frac{1}{4}\delta Z_V V_{a\mu\nu} V_a^{\mu\nu}+\frac{1}{2}\delta \{M_V^2\}{V}_{a\mu}V_a^{\mu}
 -\frac{1}{4}\delta \{g^2\}f_{abc}f_{ade}V_{b\mu}V_{c\nu} V_d^{\mu}V_e^{\nu}-\delta \{g\}f_{abc}\partial_{\mu}V_{a\nu} V_b^{\mu}V_c^{\nu}\\
&+i\delta Z_\Psi\bar{\Psi}_a \gamma^{\mu}\partial_{\mu}\Psi_a-\delta \{m\}\bar{\Psi}_a\Psi_a
+i\delta \{\text{G}_\text{F}\}f_{abc}\bar{\Psi}_{a}\gamma^{\mu}\Psi_{b} V_{c\mu}+\delta \{\text{G}_\text{D}\}d_{abc} \bar{\Psi}_{a}\gamma^{\mu}\Psi_{b}V_{c\mu}.
\end{aligned}
\end{equation}
   By comparing the expression for $\delta g$ obtained from the $VVV$- and $VVVV$-vertex functions, we can
test whether there are relations among the coupling constants $g$, $\text{G}_\text{F}$, and $\text{G}_\text{D}$.
   The ``worst case'' would be if all couplings were independent.
   A different scenario is a type of universality, relating $g$ to a linear combination of
$\text{G}_\text{F}$ and $\text{G}_\text{D}$.
   Finally, all couplings could be expressible in terms of a single parameter.

\section{Conclusion}

   At the classical level, the standard lowest-order SU(3)-invariant Lagrangian involving three independent coupling
constants $g$, $\text{G}_\text{F}$, and $\text{G}_\text{D}$ is a self-consistent starting point.
   This was explicitly shown using Dirac's method.
   Whether the requirement of perturbative renormalizability implies additional constraints among
the couplings remains to be seen.

\end{document}